# Chalcogen Assisted Enhanced Atomic Orbital Interaction at TMDs – Metal Interface & Chalcogen Passivation of TMD Channel For Overall Performance Boost of 2D TMD FETs

Ansh[1], Jeevesh Kumar[1], Gaurav Sheoran[1], Harsha B. Variar[1], Ravi K. Mishra[2], Hemanjaneyulu Kuruva[1], Adil Meersha[1], Abhishek Mishra[1], Srinivasan Raghavan[2] and Mayank Shrivastava[1]
[1]Department of Electronic Systems Engineering and [2]Center for Nanoscience & Engineering
Indian Institute of Science, Bangalore, India email: mayank@iisc.ac.in

***Abstract —*** Metal-semiconductor interface is a bottleneck for efficient transport of charge carriers through Transition Metal Dichalcogenide (TMD) based field-effect transistors (FETs). Injection of charge carriers across such interfaces is mostly limited by Schottky barrier at the contacts which must be reduced to achieve highly efficient contacts for carrier injection into the channel. Here we introduce a universal approach involving dry chemistry to enhance atomic orbital interaction between various TMDs ($MoS_2$, $WS_2$, $MoSe_2$ and $WSe_2$) & metal contacts has been experimentally demonstrated. Quantum chemistry between TMDs, Chalcogens and metals has been explored using detailed atomistic (DFT & NEGF) simulations, which is then verified using Raman, PL and XPS investigations. Atomistic investigations revealed lower contact resistance due to enhanced orbital interaction and unique physics of charge sharing between constituent atoms in TMDs with introduced Chalcogen atoms which is subsequently validated through experiments. Besides contact engineering, which lowered contact resistance by 72, 86, 1.8, 13 times in $MoS_2$, $WS_2$, $MoSe_2$ and $WSe_2$ respectively, a novel approach to cure / passivate dangling bonds present at the 2D TMD channel surface has been demonstrated. While the contact engineering improved the ON-state performance ($I_{ON}$, $g_m$, $\mu$ & $R_{ON}$) of 2D TMD FETs by orders of magnitude, Chalcogen based channel passivation was found to improve gate control ($I_{OFF}$, SS, & $V_{TH}$) significantly. This resulted in an overall performance boost. The engineered TMD FETs were shown to have performance on par with best reported till date.

## I. Introduction

Growth of semiconductor industry is driven by Moore's law[1] which intends to improve the efficiency of electronic gadgets in terms of speed and compactness by 2×, every 1.5 years. This is achieved by aggressive channel length scaling of Silicon MOSFETs. On the other hand, channel length scaling leads to short channel effects (SCE) like drain induced source barrier lowering and threshold voltage roll-off due to compromised gate control over channel. This results into higher source-to-drain leakage current, higher subthreshold slope and lower noise margins, which eventually increases the static power loss across the VLSI system. To mitigate SCE, devices like FinFETs[3, 4, 5] Multi-gate FET [2, 3, 6, 7], Ultra-thin body (UTB) FETs[6, 8] and Tunnel FETs (TFETs)[9] have been proposed, which offer improved gate control and better SCE immunity. The key in most of the ultra-scaled FET concepts is to reduce the channel thickness as the channel length is scaled down. However, scaling channel thickness beyond 5nm leads to mobility degradation and threshold voltage instability due to quantum confinement and surface dangling bonds, which leads to performance roll off. Atomically thin layers of 2D semiconductors like Transition Metal Dichalcogenides (TMDs)[10-15] on the other hand offer better gate control due to lack of dangling bonds perpendicular to their basal plane, as well as missing quantum issues when compared to bulk semiconductors. This makes 2D TMDs promising candidates for short channel FETs. While a decent amount of work has been reported by various authors on improving performance of 2D TMDs like $MoS_2$[13, 22, 24, 25, 26, 34], $MoSe_2$[16, 17], $WS_2$[18, 19] and $WSe_2$[20, 21], they still suffer from high contact resistance, low ON state current, depletion mode operation and poor sub-threshold slope.

Techniques reported earlier to improve device performance were often parameter-specific and while they resulted in striking improvement in the target parameter, the overall transistor behavior was often compromised, or the technique fails to offer a scalable process. For example, in earlier works, techniques like doping by Potassium[27], PEI[28], Chloride ion[29], Benzyl Viologen[30], Methanol[31], Tetracyanoquinodimethane (TCNQ)[32], phase engineering[33] on $MoS_2$ and/or $WS_2$ have been utilized to improve ON state current. These methods however suffered from one or the other limitations like non-scalability, involvement of wet chemistry or deterioration of other figure of merit parameters. For instance, Scandium contacts have resulted in record high ON currents in $MoS_2$[15], however at the cost of significant increase in OFF current. Besides, Scandium is a highly unstable metal, which is not suitable for scalable largescale process. Therefore, it would not be an exaggeration to say that absence of a scalable or CMOS equivalent process to boost overall device behavior is a key bottleneck in the development of 2D TMD device technology. In this work, we present a scalable and universal method to improve contact as well as channel properties of 2D TMD FETs. Based on fundamental insights into the quantum chemistry of TMD/metal contacts and mechanism of molecular decomposition of Chalcogen precursor at low temperature, we introduce an $H_2S$ based dry chemistry to offer better contact and channel properties for a range of TMDs.

## II. Quantum Chemistry of TMD/Metal Contacts

To begin with, impact of Chalcogen atom at the interstitial sites between TMD-Metal interface, for various material systems - $MoS_2$/Ni, $MoSe_2$/Ni, $WS_2$/Cr and $WSe_2$/Ni, was studied using following three contact topologies: (i) defect-free interface, (ii) defected (with Chalcogen-vacancy) interface and (iii) metal-doped (with Chalcogen atom at interstitial site) interface. It is observed (Fig. 1 & 2) that having a Chalcogen atom at an interstitial site on the TMD surface results in significant spread in the Transmission coefficient and Density of States (DOS) across a wider range of energy levels. Qualitatively, it reveals enhanced carrier injection through contacts with Chalcogen atom at the interstitial site. The effect of chalcogen atom on the interface is quantified through contact current calculations which directly are related to contact resistance. Figure 3 shows lower interface resistance with the introduction of Chalcogen atom at the TMD-Metal interface. These computational findings predict possibility of contact improvement with Chalcogen treatment over the TMD surface. It should be noted that the findings are universal for Se and S based TMDs.

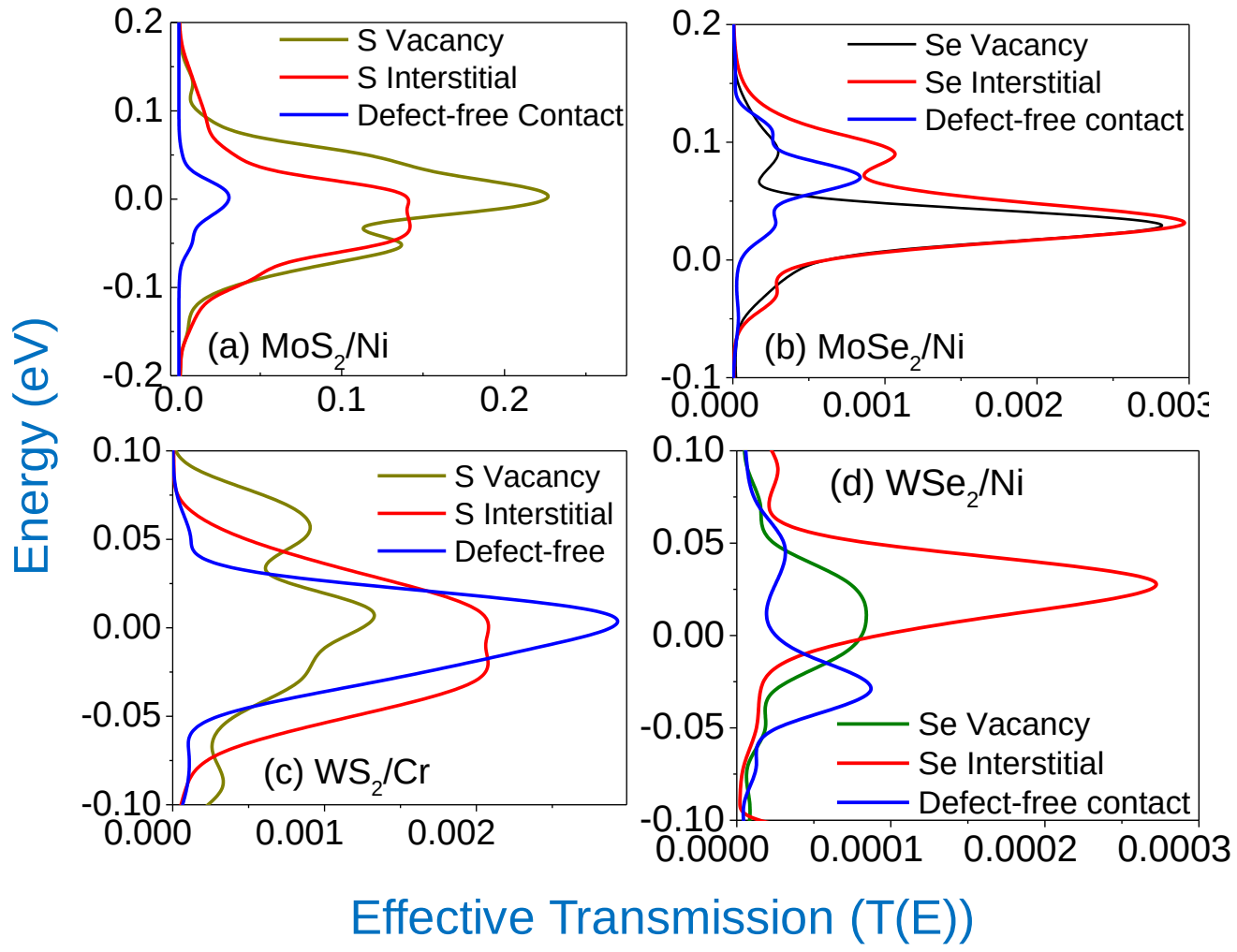

Fig. 1: Effective transmission spectrum, T(E) for various 2D TMD/metal interfaces with varying contact topology -- (a) $MoS_2$/Ni, (b) $MoSe_2$/Ni, (c) $WS_2$/Cr and (d) $WSe_2$/Ni is calculated using Extended Huckel Semi empirical method. Higher transmission peak near Fermi-level together with wider spread across energy states is observed in presence of Chalcogen interstitial/defect, when compared to standard contacts for all the interfaces. The extent of improvement was however found to be independent of TMD-Metal system.

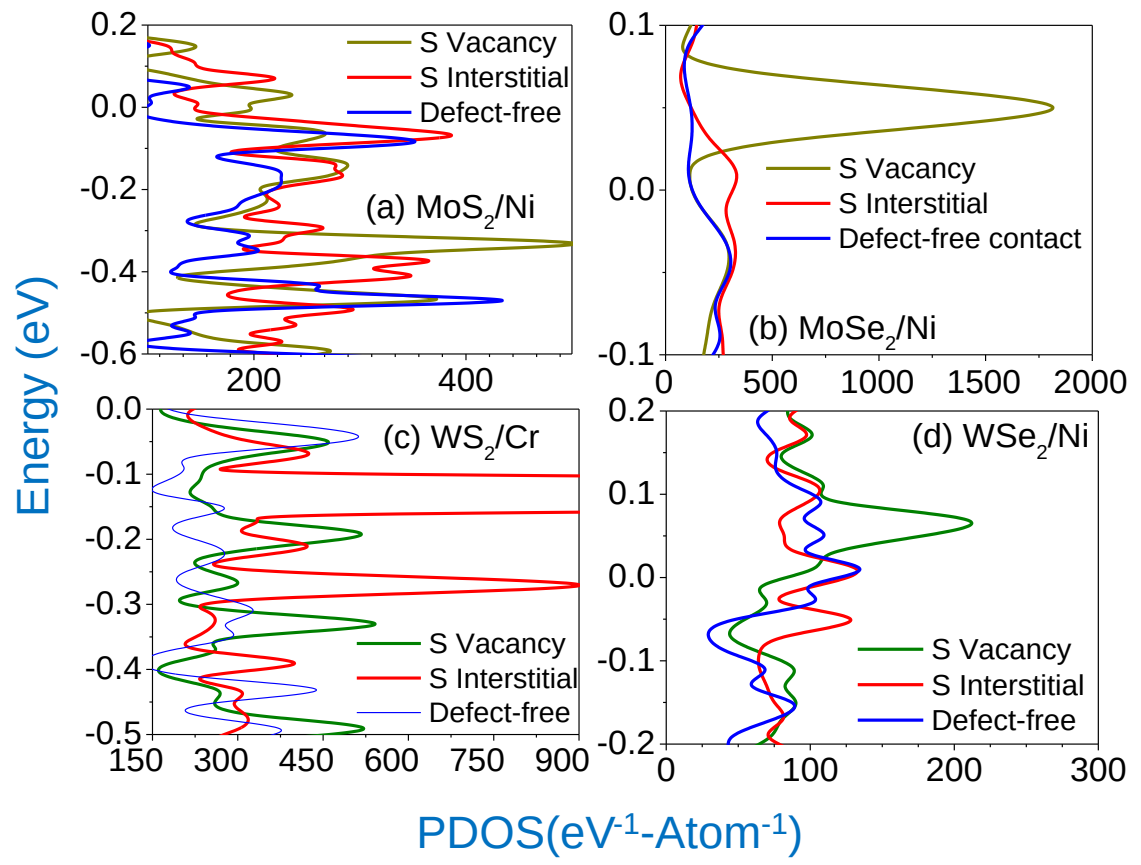

Fig. 2: Partial density of states (PDOS) using different 2D TMD/metal interfaces – (a) $MoS_2$/Ni, (b) $MoSe_2$/Ni, (c) $WS_2$/Cr and (d) WSe2/Ni – were calculated for three different topologies using Extended Huckel Semi empirical method. Enhanced PDOS values near the Fermi level are observed in Chalcogen interstitial contact topology hence, making this contact topology intriguing for further investigations.

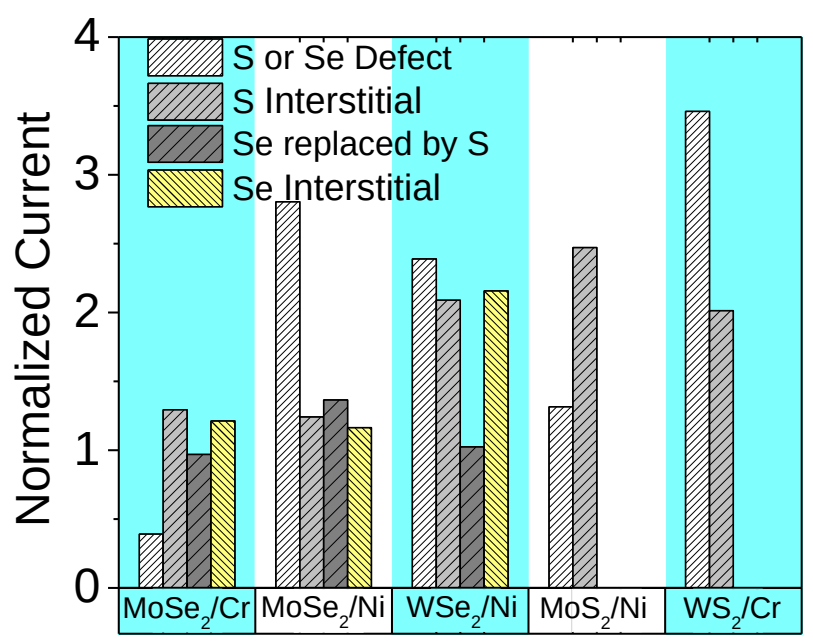

Fig. 3: Current across different TMD/metal contacts with all possible contact topologies involving Chalcogen atoms, normalized with current across standard contact. Current across the contact has a direct relation with contact resistance. Therefore, higher current implies lower contact resistance ($R_C$) in case of Chalcogen interstitial contact topologies than in case of standard contact. Physical understanding of mechanism behind lower contact resistance in Chalcogen interstitial should help in engineering the contacts at 2D TMD/metal interfaces.
.

Further Mülliken Charge Population (MCP) was extracted for the constituent atoms to understand how the metal-TMD interface is altered after introducing Chalcogen atom at various TMD-metal interfaces, as depicted in Figure 4. It is observed, that for all TMD-metal interfaces, the amount of average charge shared in the contact region (i.e. at the interface) by all the constituent atoms has increased from its reference value in the non-contacted bulk crystals. This increase was found to be highest for the topology with Chalcogen interstitial. Observation of lower contact resistance with increased charge share across atomic species at the interface implies enhanced orbital interaction between metal and TMD, which resulted in improved bonding and contact property. It is worth highlighting that the source of such an increase in the charge around all atomic species at the interface must be the semiconductor bulk and not the introduced Chalcogen atom. This can be visualized as higher electron density/concentration at the contact region, which classically can be considered as doping of the semiconductor crystal at the metal-TMD interface. The effect of such Chalcogen assisted self-doping of TMD-metal interface on its interface properties is expected to be unique for different contact metals depending on the available overlapping atomic orbitals for interaction or hybridization.

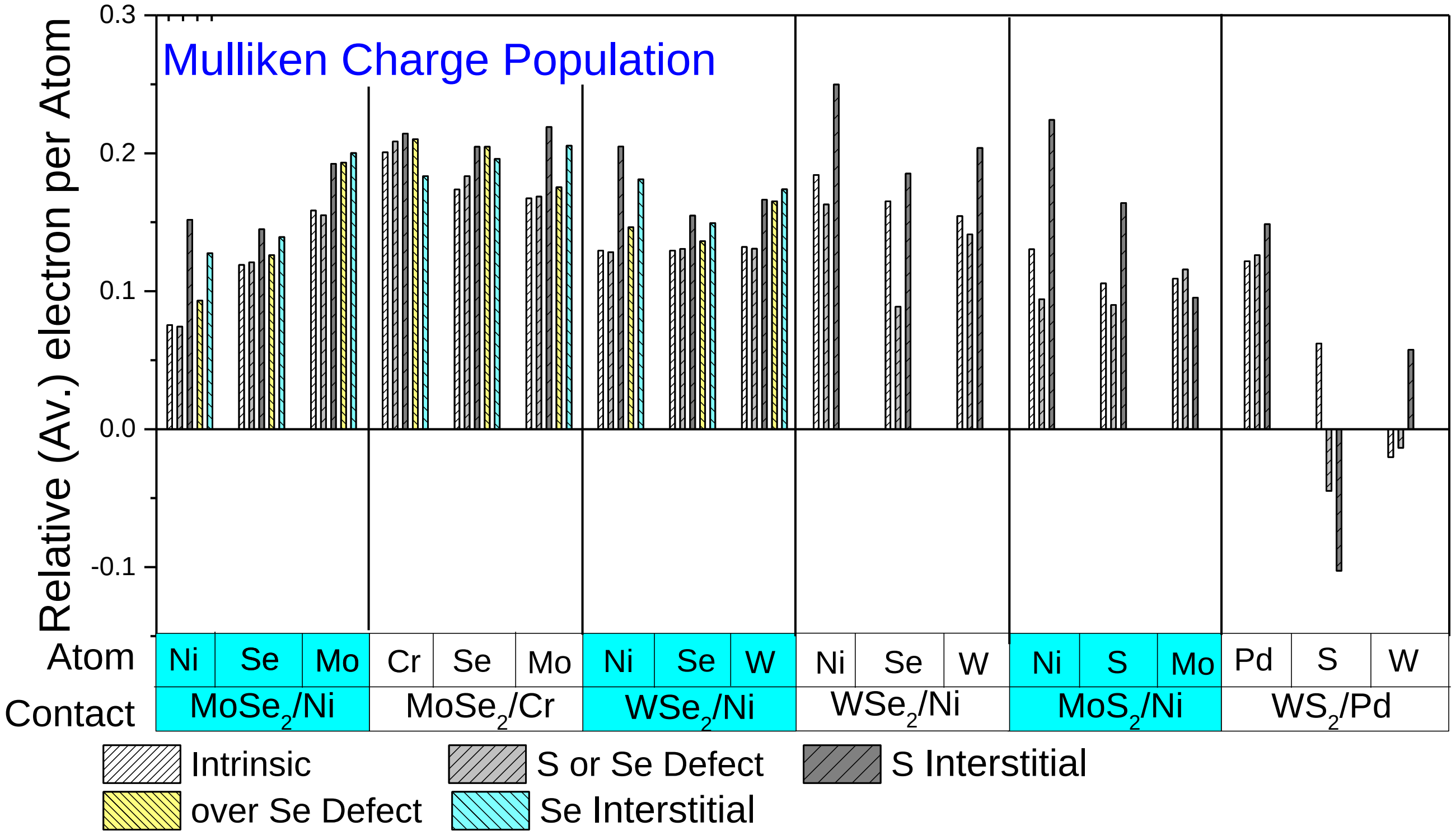

Fig. 4: Mulliken Charge Population (MCP) extracted for all the contact topologies discussed earlier. Shift in the electron cloud towards the interface atomic species is predicted from MCP.

## III. Experimental Framework

Based on computational finding, now the problem statement is to introduce Chalcogen atom at the TMD-metal interface. $H_2S$ is a widely used Sulfur source for CVD growth of S based TMDs, which partially decomposes at lower temperatures in presence of TMDs [35]. In the presence of TMD, beside satisfying S vacancies, $H_2S$ also tends to get adsorbed at TMD surface thereby forming S-S (S-Se) bond between S from $H_2S$ and S (Se) bonded to an adjacent Mo/W atom in TMD as depicted in Fig 5. Typically, complete decomposition of $H_2S$ requires higher temperatures, however, presence of a TMD surface catalyzes the decomposition reaction and $H_2S$ breaks into S. Here S stays bonded with S/Se atom at the TMD surface and $H_2$ leaves the surface after initial adsorption as discussed above[35]. In order to release S from the surface of TMD another catalytic reaction is carried out which ensures complete low temperature decomposition of $H_2S$ into S and $H_2$. This decomposition process was earlier validated by work in ref. 31. In this work, we perform only the first step of complete decomposition of $H_2S$ i.e. bonding of S on the TMD surface followed by release of $H_2$. Subsequent step to catalyze the release of S is avoided intentionally so that a contact topology with S interstitial can be achieved. Complete mechanism of low temperature partial decomposition of $H_2S$ over the TMD surface while using proposed $H_2S$ treatment and device fabrication flow is depicted in Figure 5. To uniformly introduce S doping over TMD top layer, TMD samples were treated with $H_2S$ inside a Quartz tube at 350 °C while varying $H_2S$ partial pressure. Details of the treatment are discussed in Supplementary information.

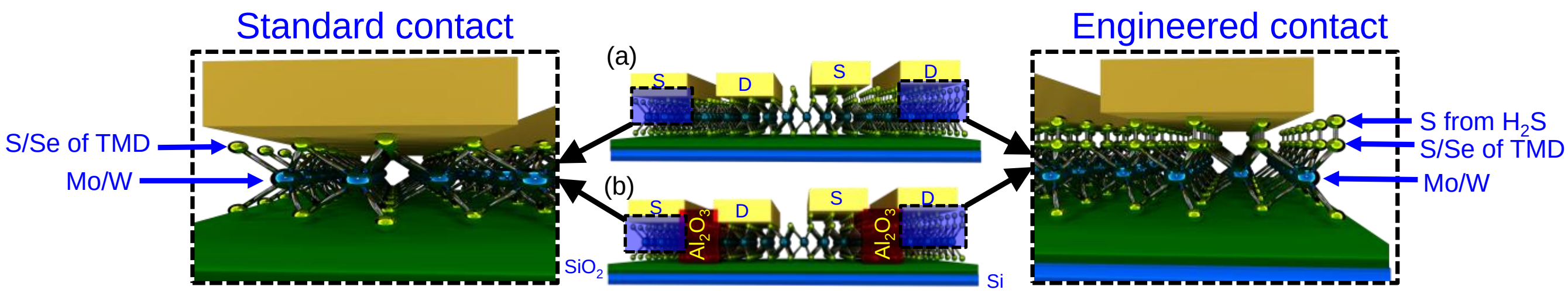

Fig. 5: Scheme representing fabricated (a) standard and contact & channel engineered and (b) standard and contact engineered transistors on the same TMD flake. Yellow spheres represent Chalcogen atoms (Sulfur and/or Selenium) whereas Blue spheres represent Transition metal atom (Molybdenum and Tungsten).

Samples were characterized using AFM, Photoluminescence, Raman spectroscopy and XPS before and after $H_2S$ treatment. AFM surface map of $MoS_2$, as depicted in Figure 6, reveals that $H_2S$ treatment leads to a smoother surface by curing surface defects present on an otherwise untreated TMD surface. Further, PL spectra shown in Figure 7 shows that the material properties in terms of bandgap and semiconducting nature remain unchanged post $H_2S$ treatment. Raman spectra depicted in Figure 8, reveals two peaks – $E_2g$ and $A_1g$ which correspond to in-plane and out-of-plane phonon modes, respectively, which are present before and after $H_2S$ exposure. This confirms that the fundamental molecular structure of the TMD remains intact post treatment. Post $H_2S$ exposure, a blue shift in $A_1g$ peak is observed, which is attributed to increased electron phonon interaction at the TMD surface[37]. This is due to shift in the electron cloud towards the surface in presence of added S atoms. It should be noted however that, since S is a highly electronegative element, this electron cloud is confined to the S atoms. This also results in a positive shift in $V_T$, which will be disclosed in subsequent sections. Raman spectra further shows lower full width half maximum (FWHM) after treatment which signifies lesser defects in the treated sample. It should be noted that most of the defects in TMDs are Chalcogen vacancies on the surface, which was found to reduced post $H_2S$ treatment. The same is further validated using XPS, as depicted in Figure 9(a) and (b). Here Mo:3$d_{3/2}$, Mo:3$d_{5/2}$ and S:2s has resulted from reduction of +5 and +6 oxidation states of Mo which reveals relatively lower number of sulfur vacancies in the treated sample[38]. The same defect mitigation can be seen for W 4f and 5p states in $WS_2$ from figure 9(c). As shown in figure 9(a) and (b), additional S peaks are obtained which could be a result of S-S or Se-S bonds at the surface of TMDs and(or) S-Mo bonds on the surface of $MoSe_2$ which were absent in an otherwise pristine $MoSe_2$ surface. In summary, AFM (Fig. 6), PL (Fig. 7), Raman (Fig. 8) and XPS (Fig. 9) spectra captured before and after $H_2S$ assisted surface engineering confirm (i) increase in the electron concentration at the surface of TMD, (ii) passivation of S-vacancies and (iii) presence of S-S (S-Se) bonds in S (Se) based TMDs when treated with $H_2S$.

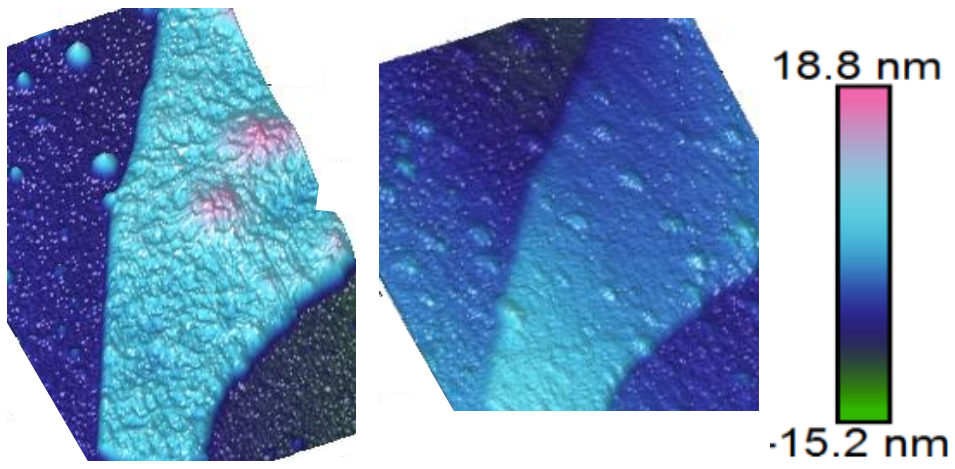

Fig. 6: AFM image of the $MoS_2$ flake before (left) and after (right) the $H_2S$ based atomic orbital overlap engineering, clearly depicts significant reduction in RMS surface roughness and removal of residues from the top.

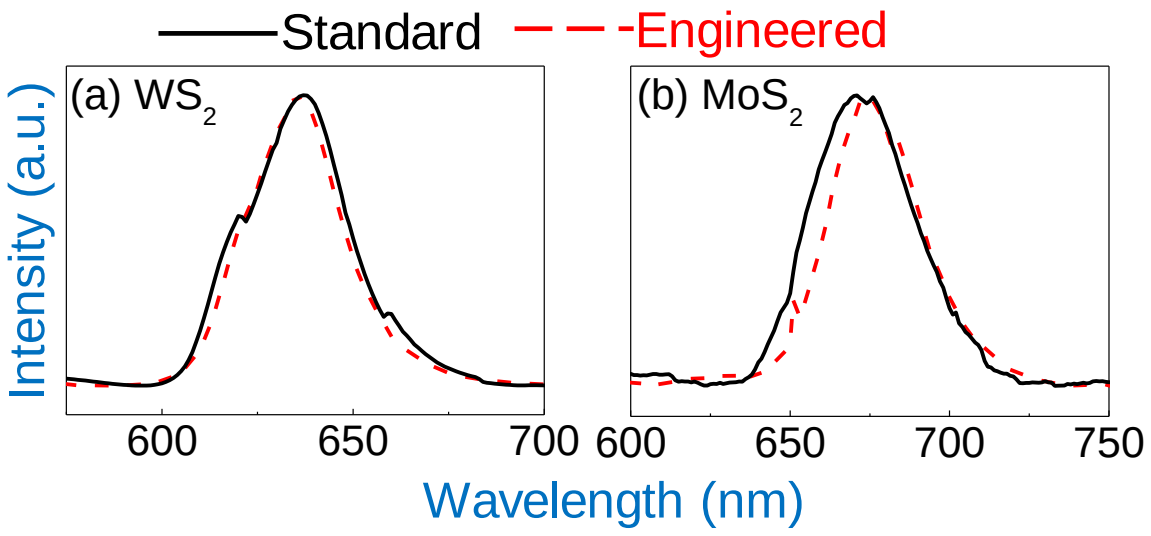

Fig. 7: PL spectra of (a) $WS_2$ and (b) $MoS_2$, before and after $H_2S$ treatment. The peaks correspond to direct bandgap (675nm/1.8eV for $MoS_2$ and 640nm/1.94eV for $WS_2$) do not show significant shift in frequency, which confirms no change in the direct bandgap of the material. This shows that the material was conserved without a noticeable change in its semiconducting properties.

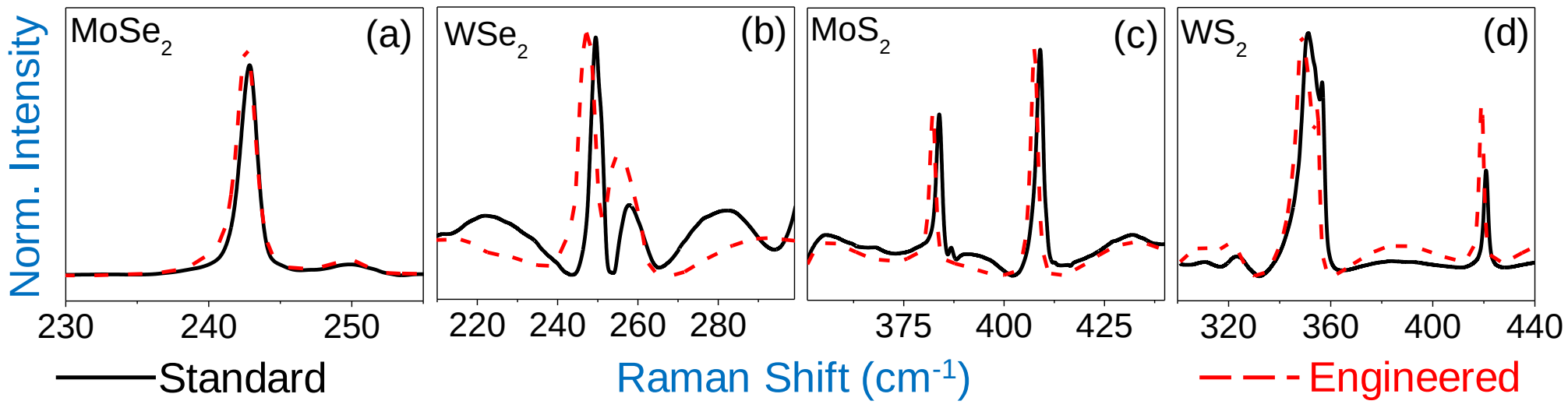

Fig. 8: (a)-(d) Raman spectra of as-exfoliated and engineered ($H_2S$ treated) $MoSe_2$, $WSe_2$, $MoS_2$ and $WS_2$ flakes. Narrowing of the Raman peaks along with lowering of shoulders close to peaks depict satisfaction of Sulfur vacancies. A red shift is consistently observed for all TMDs implying increased electron-Phonon coupling. This observation validates the anticipated (from MCP calculations) shift in electron cloud towards the surface from bulk after $H_2S$ exposure.

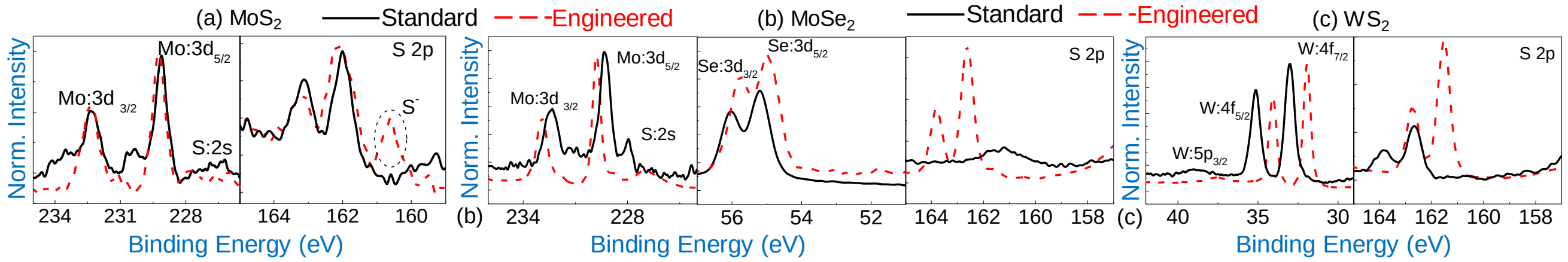

Fig. 9: (a)-(c) XPS spectra for $MoS_2$, $MoSe_2$ and $WS_2$ samples before and after $H_2S$ exposure show reduced shoulders near 3d peaks of Mo which imply reduction in S vacancy/defects and Mo dangling bonds present in the material. This validates the theory of defect curing. Presence of S in engineered $MoSe_2$ surface reveals that $H_2S$ exposure is a reliable method to introduce S atoms at the surface to enable unique charge distribution mechanism and atomic orbital overlap at the interface.

## IV. Experimental Validation and Results

Implications of such a treatment on the device thereby engineering the contact and/or channel regions to realize desired device behavior are studied by realizing back gated TMD FETs, details of which are discussed as methods. In order to capture the effect of $H_2S$ treatment on contact as well as channel, devices were exposed to $H_2S$ in the (i) contact region and (ii) channel as well as contact region. A flow that allows to independently quantify the effect of chalcogen based engineering over channel and contact was adopted. Contact engineered $MoS_2$ and $WS_2$ FETs exhibit 3× and 5× higher ON current, respectively, without affecting OFF-state current and threshold voltage when compared with standard FETs (Fig. 10a-b). On the other hand, the ON & OFF-state performance of contact engineered $MoSe_2$ FETs have improved by 2× and 5 orders, respectively (Fig. 10c). Improved ON-state current is a result of S assisted charge sharing at the TMD-metal interface. The OFF-state performance of contact engineered $MoS_2$ and $WS_2$ FETs are unaffected by $H_2S$ exposure as the channel was masked against $H_2S$ treatment thereby retaining its intrinsic channel properties. However, $H_2S$ assisted contact engineering on $MoSe_2$ FET led to a complete device turn OFF at $V_{GS}$ = -25 V which was otherwise ON for the entire range of $V_{GS}$. Such a drastic reduction of OFF-state current and enhancement mode operation is explained to be an effect of stronger interaction between S interstitial atom and electrons in its vicinity that keeps the device turned OFF for a higher $V_{GS}$. This interaction is relatively weak in $MoS_2$ and $WS_2$ FETs than in $MoSe_2$ due to difference in the atomic orbitals involved in Se based TMDs when compared to S based TMDs. Contact with channel engineered $MoS_2$ and $WS_2$ FETs exhibit 20× & 80× improvement in ON-state performance and 2 & 1 order reduction in OFF-state leakage, respectively. Moreover, $MoSe_2$ FETs confirm to have large positive $V_T$ and therefore exhibit 5 orders of magnitude lower OFF-state current when compared to standard FETs (Fig. 10d-f).

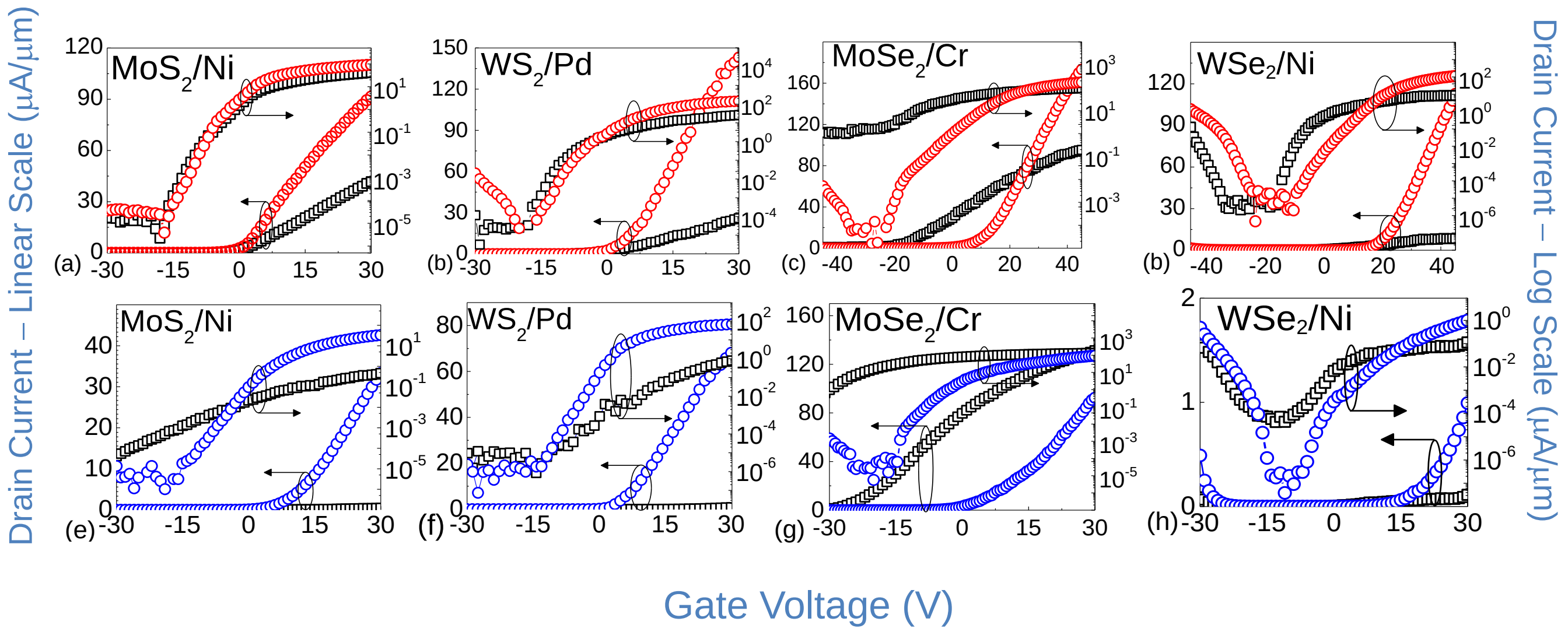

Fig. 10: (a)-(c) Contact engineered $MoS_2$, $WS_2$ and $MoSe_2$ FETs. ($L_G$ = 1 μm, $V_{DS}$ = 4 V, Gate dielectric = 90nm, $SiO_2$). (d)-(f) Channel with contact engineered $MoS_2$, $WS_2$ and $MoSe_2$ FETs. ($L_G$ = 1 μm, $V_{DS}$ = 1 V, Gate dielectric = 90 nm, $SiO_2$). For $WS_2$ channel with contact engineered devices $I_D$ = 240μA/μm was achieved, shown in Figure6 of Supplementary information.

*Overall Performance Boost:* Typical doping techniques reported elsewhere result in increased leakage current and suppressed gate control. In this work, however, leakage current has reduced by a significant amount leading to higher current modulation along with improved gate control as well as other performance metrics in treated FETs. In general, for all TMDs, contact engineering with $H_2S$ leads to improved contact performance and proposed channel engineering offers an improved OFF-state behavior beside improved gate control. Such a unique improvement in the overall device performance has been demonstrated for the first time. It should however be noted that the magnitude of improvement depends on TMD material and contact metal. Fig. 11 summarizes the overall performance improvement in terms of various figure of merit parameters ($I_{ON}$, $g_m$, μ, $I_{OFF}$, SS, $I_{ON}/I_{OFF}$ and $R_C$) by using the proposed contact and channel engineering technique. Significantly improved mobility, $R_C$ and transconductance are results of transparent contact and improved channel properties. It is worth mentioning that contact resistance is extracted using Y-function technique and the details of extraction are discussed in the supplementary information.

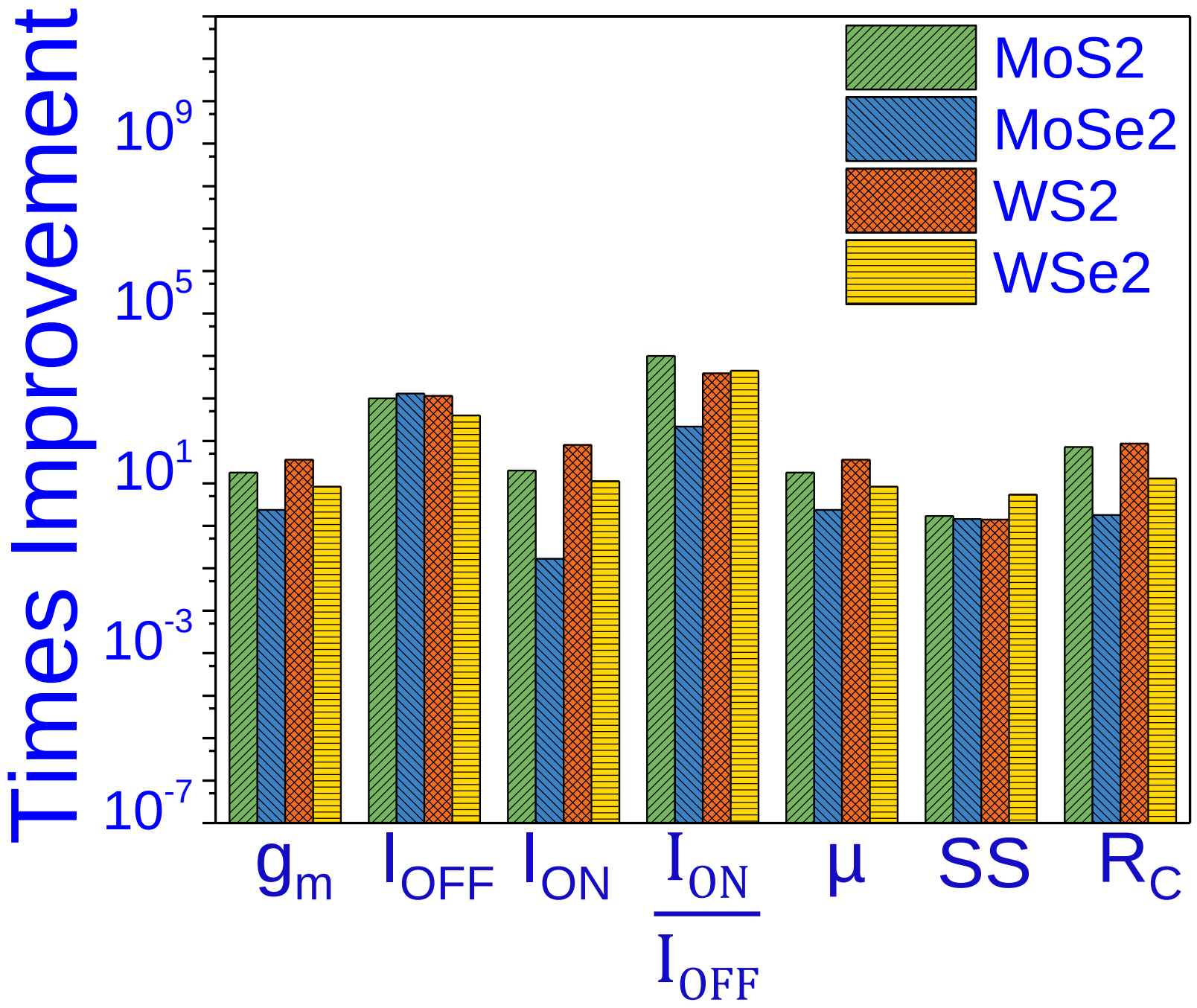

Fig. 11: Improvement in performance figure of merit parameters for various TMD FETs while using proposed contact or/and channel engineering.

In order to probe the effect of this treatment on the metal-TMD interface, Schottky Barrier Height (SBH) is extracted using the Thermionic Emission Theory[39, 40]. Figure 12 shows that SBH for $WS_2$/Pd after treatment is negative (-20 meV) unlike the SBH (46 meV) before treatment which is positive. Negative SBH implies an Ohmic contact[35]. Another important observation is that the pinning factor (S-factor) for contact and channel engineered device has reduced which reflects stronger pinning of the Fermi level in engineered samples. Strong Fermi level pinning is a result of increase in the number of surface states[40, 41]. Increase in the number of surface states is attributed to the presence of more acceptor like states within the bandgap as a result of formation of S-S bonds after the treatment. Details of extraction and analysis are in supplementary information.

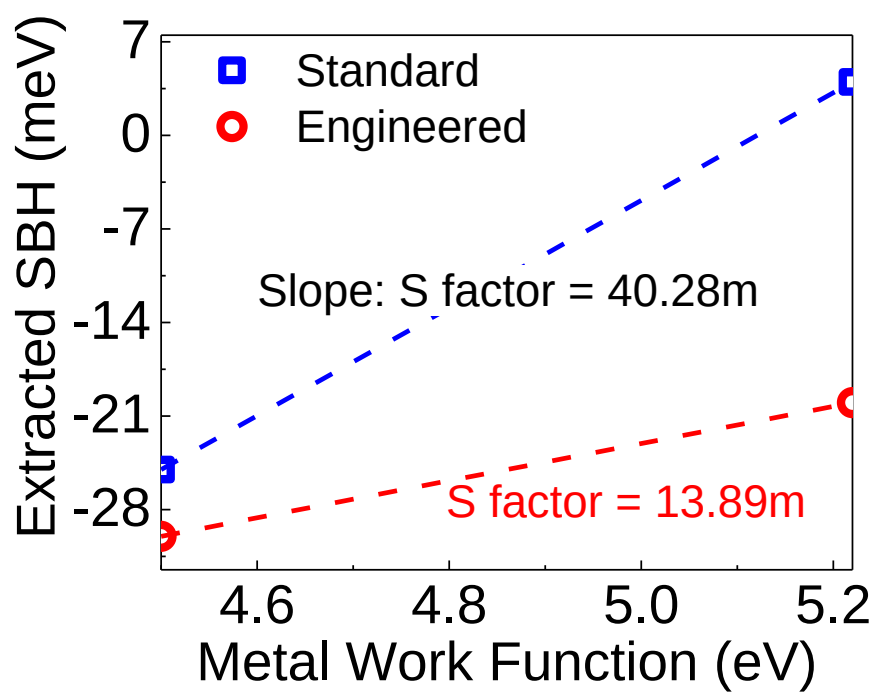

Fig: 12. Extracted Schottky barrier heights of standard and engineered devices with different contacting metals for $WS_2$ FETs. Pinning factor shows stronger pinning of the Fermi-level after $H_2S$ assisted engineering of TMDs. Negative SBH is achieved after engineering which implies presence of ohmic contact.

Presence of acceptor like states on the TMD surface is confirmed using atomistic simulations, as shown in Figure 13. Density of States (DOS) of three types of monolayer $WS_2$ structures – (i) defect-less structure, (ii) with S vacancies and (iii) with S atom as interstitial, are calculated. Defect less crystal does not have any states in the bandgap unlike other two structures where surface states appear within the bandgap. For $WS_2$ with S vacancy, surface states lie near the conduction band minimum while for $WS_2$ with S interstitial has surface states near the valence band maximum.

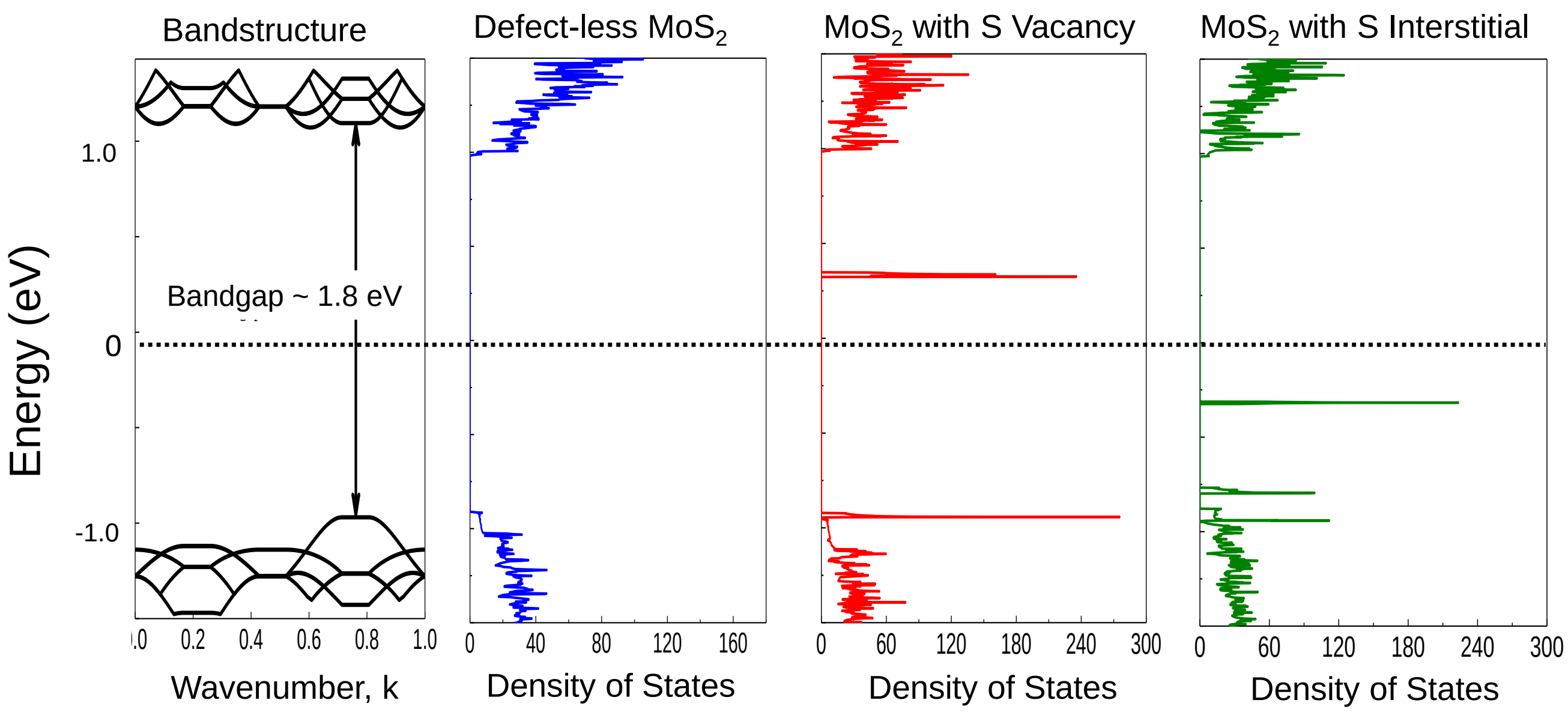

Figure 13: (a) Bandstructure of monolayer $MoS_2$ exhibiting a direct bandgap. (b) Density of states (DOS) in a defect-less $MoS_2$ monolayer. Absence of energy states within the bandgap can be observed. (c) DOS in a monolayer $MoS_2$ with S-vacancy. Presence of Donor type surface states within the bandgap is observed. (d) DOS in a monolayer $MoS_2$ with an S interstitial atom.

*Monolayer CVD TMD:* Physical insights into the unique charge sharing mechanism and its experimental validation encouraged positive implications of this process on CVD TMD monolayer so that its technological relevance can be unveiled (Fig. 14). ON-state performance of CVD monolayer $MoS_2$ FETs was found to improve by 4× and 3× for contact and contact with channel engineered FETs, respectively. Besides, the OFF-state leakage reduced by 1 order of magnitude for both the types of engineered devices. Standard CVD monolayer $MoS_2$ FETs are limited in terms of performance due to high defect density (manifests as plateau in transfer characteristics, Fig. 14) and high contact resistance due to large bandgap. Disappearance of the plateau in transfer curve further confirms elimination of defect states because of proposed channel engineering. Finally, a noticeable shift in threshold voltage was observed, leading to first demonstration of enhancement mode CVD $MoS_2$ FETs. This is also attributed to reduction in S vacancies after channel treatment / curing of intrinsically defected CVD TMD material.

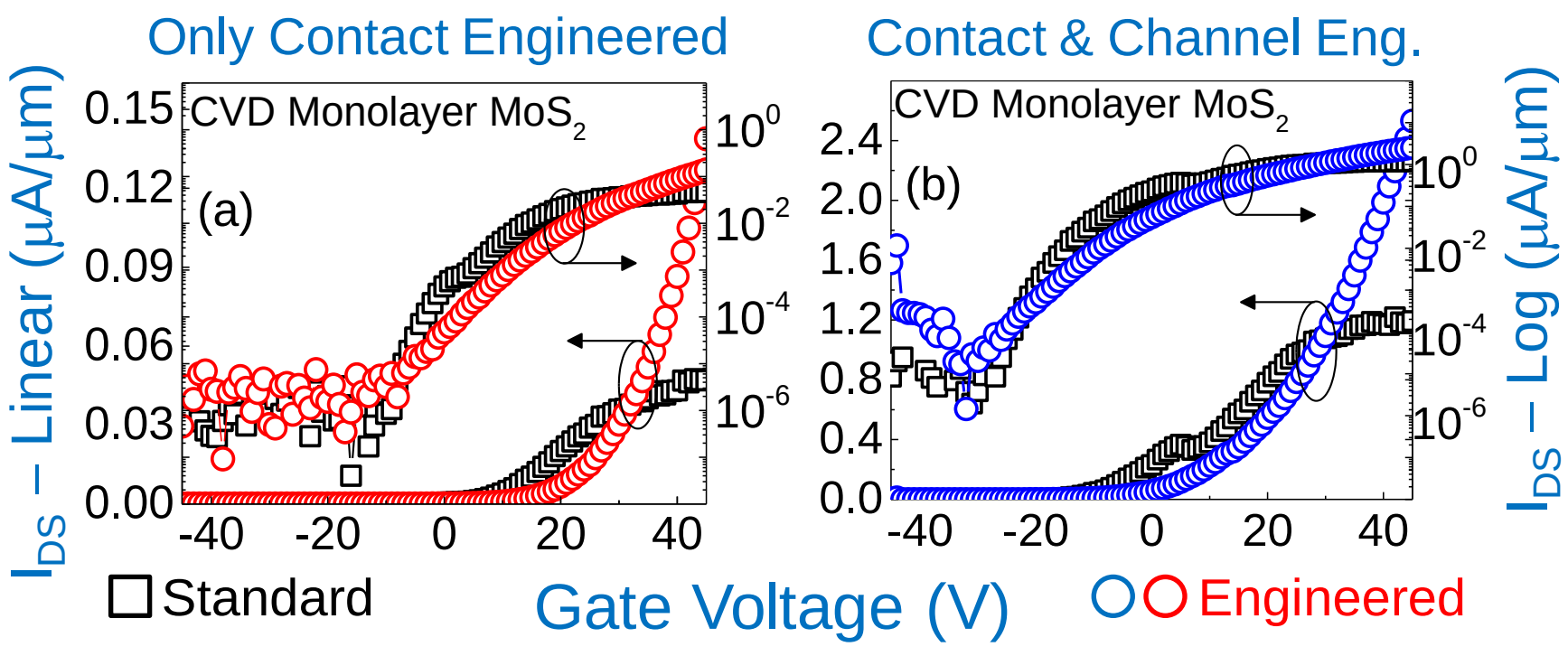

Fig. 16: (a) and (b) CVD monolayer $MoS_2$ FETs exhibit enhanced performance after $H_2S$ exposure for contact and contact with channel engineering, respectively. Defect-limited transport is enhanced by curing the defects present in CVD $MoS_2$ via channel with contact engineering while enhancing current injection through contacts. ($L_G$ = 1μm, $V_{DS}$ = 4 V, $SiO_2$ 90 nm)

This treatment has led to a significant improvement in all the device parameters of TMD based FETs. When compared to other devices in the literature, $H_2S$ treated devices – as proposed in this work – exhibit comparable or better performance when compared to best number reported in literature. This is shown in Figure 17.

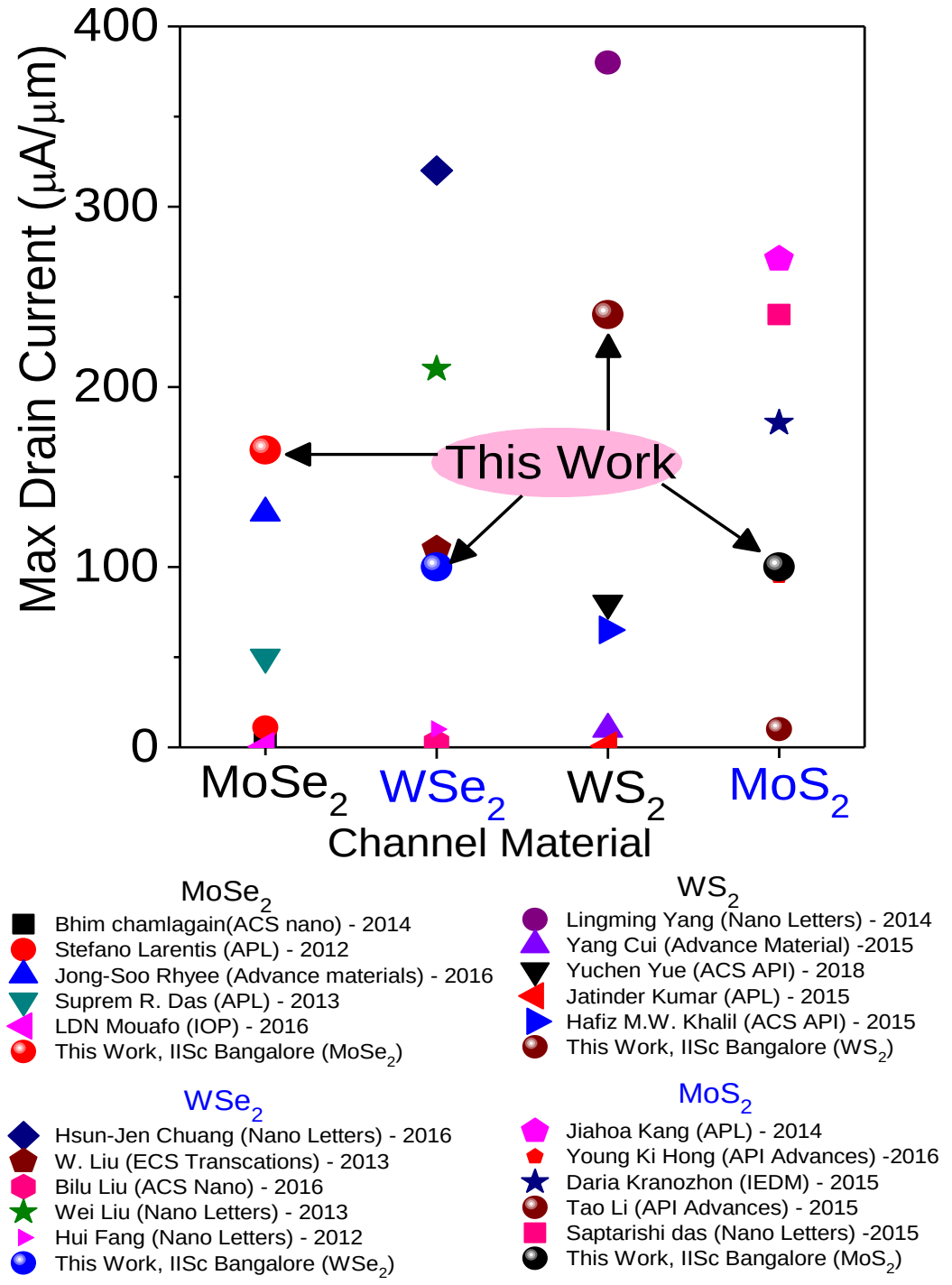

Fig. 17: ON current comparison with state-of-the-art 2D TMD FETs.

## V. Conclusion

Developed physical understanding of unique atomic orbital interaction between TMDs, Chalcogen atoms and metals, a scalable approach to engineer TMD/metal interface and TMD channel has been proposed and experimentally demonstrated for 4 different TMD materials ($MoS_2$, $WS_2$, $MoSe_2$ and $WSe_2$). S assisted enhanced orbital interaction at TMD/metal interface resulted in transparent contacts, which led to reduced metal-TMD contact resistance and improved ON-state performance. On the other hand, S assisted channel engineering, attributed to annihilation of intrinsic defects in TMD channel, has improved OFF state behavior and led to a controlled transition from depletion mode operation to enhancement mode operation of 2D FETs. Detailed understanding of the quantum chemistry involved has eased the generality of this process over S & Se based TMDs. The proposed method universally results in a significant improvement in channel and contact performance, thereby improving the overall transistor behavior, for the entire range of experimented 2D TMDs. Finally, technological relevance of the proposed scheme has been validated for monolayer CVD material.

Author contributions

Ansh came up with the idea to treat $WS_2$ with $H_2S$. MS unified the idea to other TMDs and envisaged atomic orbital overlap engineering. Device fabrication and electrical characterization were performed by Ansh ($MoS_2$, $WS_2$, $MoSe_2$, $WSe_2$), GS ($WS_2$, $WSe_2$) and HK ($MoS_2$). JK and HV performed atomistic computations and optical/electrical characterization, respectively. RM and SR contributed in setting up rector for $H_2S$ treatment and subsequently $H_2S$ treatment on all the devices. All the authors contributed in analyzing the data. Ansh, AM and MS wrote the paper.

Conflict of interests

The authors declare no competing interests.

Data availability

The data related to the findings of this work are available from the corresponding author subject to reasonable request.

**Simulation details for Device calculations:** The metal semiconductor contacts are created using Virtual Nano Lab builder in ATK. The contact metal is cleaved along the surface which gives minimum strain to the contact interface. Extended Huckel Semi Empirical method of Atomistix ToolKit and NEGF are used for the calculations. The density mesh cutoff is 45 Hartee with 10 k-points along the width and 200 k-points along the channel of the devices. Different Huckel basis sets are used for different devices along with a multigrid Poisson solver with Dirichlet boundary condition. A source–drain bias of 250 mV is applied to conduct the carrier transport analysis on 300K electron temperature.

**Simulation details for bulk calculation:** The DFT simulation is done for three different structures of the $MoS_2/WS_2$ to validate the results. For Ab-initio simulation, QuantamWise ATK simulation package has been used. The DFT calculation is performed on a 5x5 $MoS_2/WS_2$ supercell with single SV, Sulfur on the interstitial sites and on the perfect crystal. The Local Density Approximation (LDA) is used as the exchange-correlation with 7 k points sampling in the periodic direction. All the structures are optimized with 0.01eV/A force tolerance and 0.001 eV/$A^3$ stress tolerance before Band structure and DOS calculations.

**Fabrication method and $H_2S$ treatment:** The fabrication process involves mechanically exfoliating TMDs using scotch tape method. The exfoliated flakes are then transferred onto a 90nm $SiO_2/Si$ sample. Contact pads for 500 nm channel length are then patterned through e-beam lithography on few-layered flakes identified through optical microscope followed by contact metal deposition using TECPORT e-beam evaporator. Post lift-off and annealing at 250 °C, electrical characterization is done on the as-fabricated back gated FETs. After electrical characterization, samples are placed inside a chamber with 20 torr partial pressure of $H_2S$ at 350 °C, as discussed earlier. Post-treatment, contact pads are patterned followed by metal deposition and lift-off on already processed flakes (previously fabricated FET). The complete process along with the mechanism of low temperature partial decomposition of $H_2S$ is shown in Supplementary information.